\documentclass[epj]{svjour}
\usepackage{graphics}
\usepackage{graphicx}
\usepackage{amssymb,color}
\pdfoutput=1
\begin{document}
\title{Average clock times for scattering through asymmetric barriers}
\author{Bryce A. Frentz\inst{1}\thanks{\email{bfrentz@cord.edu}} \and Jos\'{e} T. Lunardi\inst{2}\thanks{\email{jttlunardi@uepg.br}} \and Luiz A. Manzoni\inst{1}\thanks{\email{manzoni@cord.edu}}
}                     
%
%
\institute{Department of Physics, Concordia
College, 901 8th St. S., Moorhead, MN 56562, USA \and Departamento de Matem\'atica e Estat\'{\i}stica,
Universidade Estadual de Ponta Grossa. Avenida General Carlos
Cavalcanti, 4748. Cep 84030-000, Ponta Grossa, PR, Brazil}
\date{Received:  / Revised version: }
%
\abstract{
The reflection and transmission Salecker-Wigner-Peres clock times averaged over the post-selected reflected and transmitted sub-ensembles, respectively, are investigated for the one dimensional scattering of a localized wave packet through an asymmetric barrier. The dwell time averaged over the same post-selected sub-ensembles is also considered. The emergence of negative average reflection times is examined and we show that while the average over the reflected sub-ensemble eliminates the negative peaks at resonance for the clock time, it still allows negative values for transparent barriers. The saturation of the average times with the barrier width (Hartman effect) is also addressed.
} 
\maketitle
%
\section{Introduction}

The search for a sensible definition of quantum tunneling times is one of the most enduring problems in quantum mechanics, despite numerous efforts in the last few decades (see, \emph{e.g.,} \cite{Win06a,Win06b,LMa94,HSt89,LMa07,DLL11,KCN09,SAk11,Gro13,CJo13}  and references therein). The difficulty in obtaining a time scale for the tunneling problem lies in the well-known impossibility of defining a self-adjoint time operator canonically conjugated to the (bounded from below) Hamiltonian. Attempts to use a ``tempus" operator that is canonically conjugated to the Hamiltonian but not given by the time evolution of the system, have lead to complex stationary times \cite{KAN94,KIG01} (also see \cite{Olk09} for a review of methods attempting to define time operators).

A more common approach to this problem is to obtain operational definitions of a parameter with the dimensions of time and investigate its properties. These attempts have rendered several definitions of time that, while useful in specific situations, are generally not considered the definitive answer to the question of how long it takes for a particle to tunnel through a potential barrier.
Among the most common (and useful) stationary time scales considered in the literature are the much studied phase time \cite{Wig55,Har62}, dwell time \cite{Smi60,But83}, Larmor time \cite{But83,Baz67,Rib67} (also see \cite{KMa12,Par13}) and the Salecker-Wigner-Peres (SWP) clock times \cite{SWi58,Per80,CLM09,Par09}.

Recently, the SWP clock has been reconsidered and it was shown that, contrary to previous claims in the literature, it can be directly applied to interacting particles (\emph{i.e.,} without the need for ``calibration'' \cite{Lea93,LMc94}) provided that one follows Peres' \cite{Per80}  approach to associate the clock's hand with the peak of the clock's wave function \cite{CLM09}. In particular, this approach allowed the derivation, directly from Schr\"{o}dinger's equation, of a relationship between the dwell time and the SWP clock reflection and transmission times without an interference term \cite{CLM09}, thus ending the controversy about the compatibility of such relationship with standard quantum mechanics (see, \emph{e.g.,} \cite{Win06a,Win06b,LMa94,HSt89} and references therein).

The SWP clock has also been used to treat the scattering of a wave packet by a potential and to obtain, by making use of a post-selection of the final state, a definition for an average traversal (reflection) time \cite{LMN11}. It was shown that, for wave packets of finite width, such average transmission time does not saturate in the limit of opaque barriers for symmetric potentials, which is an important result associated with the Hartman effect \cite{Har62}. One must notice, however, that the absence of saturation is not enough to prevent apparent superluminal speeds in the relativistic case \cite{LMNP11} -- but, as pointed by Davies \cite{Dav05}, the SWP clock times can be interpreted as weak values \cite{AAV88,AhV90}, and such results are not unexpected for weak values when they are associated with small probabilities.

The results in \cite{LMN11,LMNP11} were obtained considering \emph{symmetric} potentials, in which case the \emph{average} transmission (reflection) SWP times coincide with the average of the dwell time over the transmitted (reflected) sub-ensemble \cite{LMN11} (as a consequence of the identity between the stationary SWP and dwell times for such potentials). Therefore, in order to better understand the properties of the \emph{average} SWP clock times introduced in \cite{LMN11}, it is necessary to analyze a situation in which they differ from the dwell time \emph{averaged over the corresponding sub-ensemble}, such as for \emph{asymmetric} potentials.

Tunneling through asymmetric potentials has important applications in semiconductor heterostructures (see \cite{Dra99} and references therein) and it has been investigated due to the possibility of negative reflection times. In this context, the authors of \cite{LSp06,CLi05} show that the asymmetry of the potential can give rise to negative reflection \emph{phase} times in the stationary regime (see also \cite{Lon01} for an investigation involving asymmetric photonic band gaps). A \emph{partial} analysis of the stationary SWP clock times for an asymmetric barrier appeared in \cite{MPo03}, in which the authors considered a comparison between phase, dwell and the B\"uttiker (or Larmor) times -- the stationary SWP clock time coincides with one of Buttiker's characteristic times (namely, $\tau_y$ in \cite{MPo03}).

In this work, we extend the approach introduced in \cite{LMN11} and investigate the properties of the \emph{average} reflection and transmission SWP times for a non-relativistic particle tunneling through an \emph{asymmetric} potential barrier. The particle is represented by a gaussian wave packet (in a fully \emph{time-dependent} treatment), and for comparison we also consider the average of the dwell time over the post-selected transmitted and reflected sub-ensembles.

This paper is organized as follows. In Section \ref{clock} we briefly review the essentials of the SWP clock to measure reflection and transmission times for a wave packet scattered by a potential barrier, introducing the modifications necessary to deal with the asymmetric potential that interest us here. Section \ref{tun} presents our results for the average clock times and the comparison with the dwell time averaged over a final sub-ensemble. Finally, Section \ref{concl} is reserved to our concluding remarks.


\section{SWP clock and average reflection and transmission times}
\label{clock}

The Salecker-Wigner-Peres clock \cite{SWi58,Per80} is essentially an external quantum rotor coupled to a particle, which measures the time interval spent by the particle in a certain region of interest. This is accomplished by introducing an interaction between the clock and the particle given by $H_{\rm int}= \mathcal{P}(z) \left( - i \hbar \omega \frac{\partial}{\partial \theta}\right)$,  where $0 \leq \theta < 2\pi$ and $\omega \equiv 2\pi/(N\tau)$. The parameter $\tau$ corresponds to the clock's resolution and $ N = 2 j+1$ is the Hilbert's space dimension ($j$ is a non-negative integer or half-integer); $\mathcal{P}(z) = 1 $ if $z \in (z_1,z_2)$ and zero otherwise. It can be shown \cite{Per80} that the effect of this coupling to the stationary state of a particle with energy $E$ interacting (in one dimension) with a potential $V(z)$ is the addition of a potential barrier $\mathcal{V}_m = m\hbar \omega$ in the region $z \in (z_1, z_2)$, with $m = -j, \ldots ,+j$.

Restricting ourselves to potentials $V(z)$ with constant (but not necessarily equal) values in the regions $\left(-\infty,z_1\right)$, $\left(z_1,z_2\right)$, and $\left(z_2,\infty\right)$, which will be named respectively as regions I, II and III, and assuming the particle to be incident from the left of $z_1$, the solution of the time-independent Schr\"{o}dinger equation with potential $V^{(m)}(z) \equiv V(z) + \mathcal{V}_m\mathcal{P}(z)$ can written as
\begin{eqnarray} \label{wf}
\psi_I^{(m)}(z) &=& e^{ik_1 z} + B^{(m)}(\mathbf{k}) e^{-ik_1 z}\; ; \nonumber \\
\psi_{III}(z) &=& C^{(m)}(\mathbf{k}) e^{ik_3 z}\; ,
\end{eqnarray}
where $\psi_I$ and $\psi_{III}$ indicate the wave function in regions I and III, respectively. In (\ref{wf}) we defined ${\mathbf k} \equiv (k_1, k_2, k_3)$, with $k_1$, $k_2$ and $k_3$ being the respective wave numbers in regions I, II, and III.

The transmission time measured by the SWP clock in the stationary case is then \cite{Per80,CLM09}
\begin{equation} \label{tt}
t_c^T(\mathbf{k}) = -\hbar \left( \frac{\partial }{\partial\mathcal{V}_m}\varphi_T^{(m)}(\mathbf{k})\right)_{\mathcal{V}_m=0}\, ,
\end{equation}
where $\varphi_T^{(m)}$ is the phase of the amplitude $C^{(m)}(\mathbf{k})$ in the presence of the perturbation. A similar expression is obtained for the stationary reflection time by using the phase of the reflection amplitude $B^{(m)}(\mathbf{k})$.

In \cite{CLM09} it was shown that, as a consequence of Schr\"odinger equation, the stationary SWP clock times and the dwell time ($\tau_D$) satisfy the following relationship
\begin{equation} \label{rel}
 \tau_D (\mathbf{k}) = T(\mathbf{k}) t_c^T(\mathbf{k}) +
 R(\mathbf{k}) t_c^R(\mathbf{k})\, ,
\end{equation}
which will prove useful in what follows. In this expression $T(\mathbf{k})$ is the weak perturbation limit ($\mathcal{V}_m \rightarrow 0$) of $T^{(m)}(\mathbf{k})=\frac{k_3}{k_1}\left|C^{(m)}(\mathbf{k})\right|^2$, which is the transmission \emph{coefficient} in the presence of the clock's perturbation. In the same way, $R=\lim_{\mathcal{V}_m \rightarrow 0}R^{(m)}$, with $R^{(m)}$ being the reflection coefficient in the presence of the clock \footnote{Notice that in \cite{CLM09}, $T$ and $R$ refer to transmission and reflection \emph{amplitudes} instead of coefficients as here. This slight change in notation is convenient in order that (\ref{rel}) applies immediately also to asymmetric potentials.}.

In this work we will be mainly concerned with the average tunneling times for localized incident particles represented by a wave packet of the form
$$
\Phi (z, t) = \int\frac{dk_1}{\sqrt{2\pi}} A(k_1) e^{i(k_1 z-E(k_1)t/ \hbar)}.
$$
The initial wave function representing the particle-clock system is given by $\Psi_{\rm in}(z, t, \theta) = \Phi (z, t)v_0(\theta )$, where the clock is chosen to be in a state $v_0(\theta )$  strongly peaked at $\theta = 0$ \cite{Per80}). The wave function transmitted to the right of the potential will have the clock and the particle in an entangled state which, assuming a weak coupling between the clock and the potential, is given by \cite{LMN11}
$$
\Psi_{\rm tr} (z, t, \theta) \! = \!\! \int \!\!\!\frac{dk_1}{\sqrt{2\pi}} A(k_1) C({\mathbf k})e^{i(k_3 z-\frac{E(k_1\! )t}{ \hbar}\! ) }v_0(\theta -\omega t_c^T({\mathbf k}))\, .
$$
The \emph{average transmission time} can be obtained by post-selecting $\Psi_{\rm tr}$ as the final state, in addition to the above pre-selection of the initial state, tracing out the the particle's degrees of freedom long after the interaction has occurred and, finally, taking the expectation value of the operator giving the peak of the clock's eigenfunctions (for details see \cite{LMN11}). The resulting average transmission time is
\begin{equation}\label{av-tr}
\langle t_c^T \rangle = \int \frac{dk_1}{\sqrt{2 \pi}}  \, \rho_T({\mathbf k}) \, t_c^T({\mathbf k})\, ,
\end{equation}
where $\rho_T({\mathbf k})$ is the probability density that a particle with wave number $k_1$ be found in the ensemble of transmitted particles, which is given by
\begin{equation} \label{dt}
\rho_T({\mathbf k})=\frac{\left| A(k_1)\right|^2 T({\mathbf k})}{\int \frac{dk_1}{\sqrt{2 \pi}} \left| A(k_1)\right|^2 T({\mathbf k}) }\, .
\end{equation}
The corresponding expression for the average \emph{reflection} time can be obtained through a similar procedure, in which one post-select the reflected asymptotic sub-ensemble, and the probability density that the particle of wave number $k_1$ be found in this ensemble is obtained from (\ref{dt}) by making the change $T\rightarrow R$.

It should be noticed that an analysis of the Larmor clock for wave packets leads to a result similar to (\ref{av-tr}) \cite{FHa88} as one of the possible time scales. Similarly, (\ref{av-tr}) is the real part of the \emph{complex} average time obtained in \cite{LAe89}. Other formalisms exist leading to complex stationary times whose real part coincides with (\ref{tt}), such as the one presented by Fertig in \cite{Fer} (in fact, the complex time scale obtained in \cite{Fer} is the same introduced in \cite{SBa}) -- these approaches presumably would also lead to (\ref{av-tr}) as the real part of a complex average time when applied to wave packets. The advantage of the SWP clock formalism when compared with those is that it provides a simple way to obtain a \emph{unique} (\emph{real}) average time scale.

It is important to mention that other approaches exist in the literature to treat the tunneling of wave packets through a potential barrier that lead to average times not directly related to the above mentioned averages. In \cite{HFF87} the authors formally develop the center-of-mass clock, which to leading order leads to the phase time for a wave packet narrow in the $k$-space. A method developed in \cite{GPo93,GOR95} by using Green functions and which renders a \emph{complex stationary time} similar to B\"{u}ttiker's (but with sensitivity to the energy $E$ rather than to the potential) was extended in \cite{DOG06} to treat wave packets, with the result of obtaining an average of the phase time. The latter result is similar to the results obtained in \cite{FHa88}, with the main difference that in \cite{DOG06} the authors adopt a cutoff in energy at $V_0$ -- such a cutoff poses difficulties with respect to the localization of the tunneling particle \cite{LMN11,FHa88,HFF87}. For an analysis of the phase time averaged over the transmitted wave packet for \emph{symmetric} potentials, see \cite{BSM94} (we know of no such analysis for \emph{asymmetric} potentials).

As it is well-known, for \emph{symmetric potentials} the \emph{stationary} SWP transmission and reflection times coincide \cite{But83,FHa88}, and from  (\ref{rel}) both these times coincide with the dwell time. As a consequence, for symmetric potentials the above average transmitted (reflected) time is just the dwell time averaged over the transmitted (reflected) sub-ensemble \cite{LMN11}. However, the fact that for \emph{asymmetric} potentials the stationary SWP clock times are different from the dwell time \cite{FHa88} has important consequences for the sub-ensemble averages above defined. Thus, in order to fully understand the properties of the \emph{average} SWP clock times, it is important to investigate how they deviate from an \emph{average} of the dwell time over the same sub-ensembles, namely
\begin{equation}\label{av-d}
\langle \tau_D\rangle_{T(R)} = \int \frac{dk_1}{\sqrt{2 \pi}}  \, \rho_{T(R)}({\mathbf k}) \, \tau_D({\mathbf k})\, .
\end{equation}
For \emph{symmetric} potentials these \emph{average} dwell times coincide with the average times defined in (\ref{av-tr})-(\ref{dt}) (and the corresponding for the reflected sub-ensemble), but in the case considered here (asymmetric potentials) they have different properties (see next section). In particular, the average times defined in (\ref{av-d}) are also good candidates to be interpreted as transmission (reflection) times, since they clearly distinguish between the transmitted and reflected particles through the post-selection of the final state.

Ensemble averages similar to (\ref{av-tr})--(\ref{av-d}) can, in principle, be obtained for any well defined stationary time scale (see, e.g., \cite{FHa88,LAe89,BSM94,POl}). Finally, it is worth noticing that in the limit of spatially very wide wave packets the averages (\ref{av-tr})-(\ref{av-d}) tend to the corresponding stationary times.


\section{Average times for asymmetric barriers}
\label{tun}

\begin{figure}
\begin{center}
\includegraphics[width=7cm,height=4.0cm]{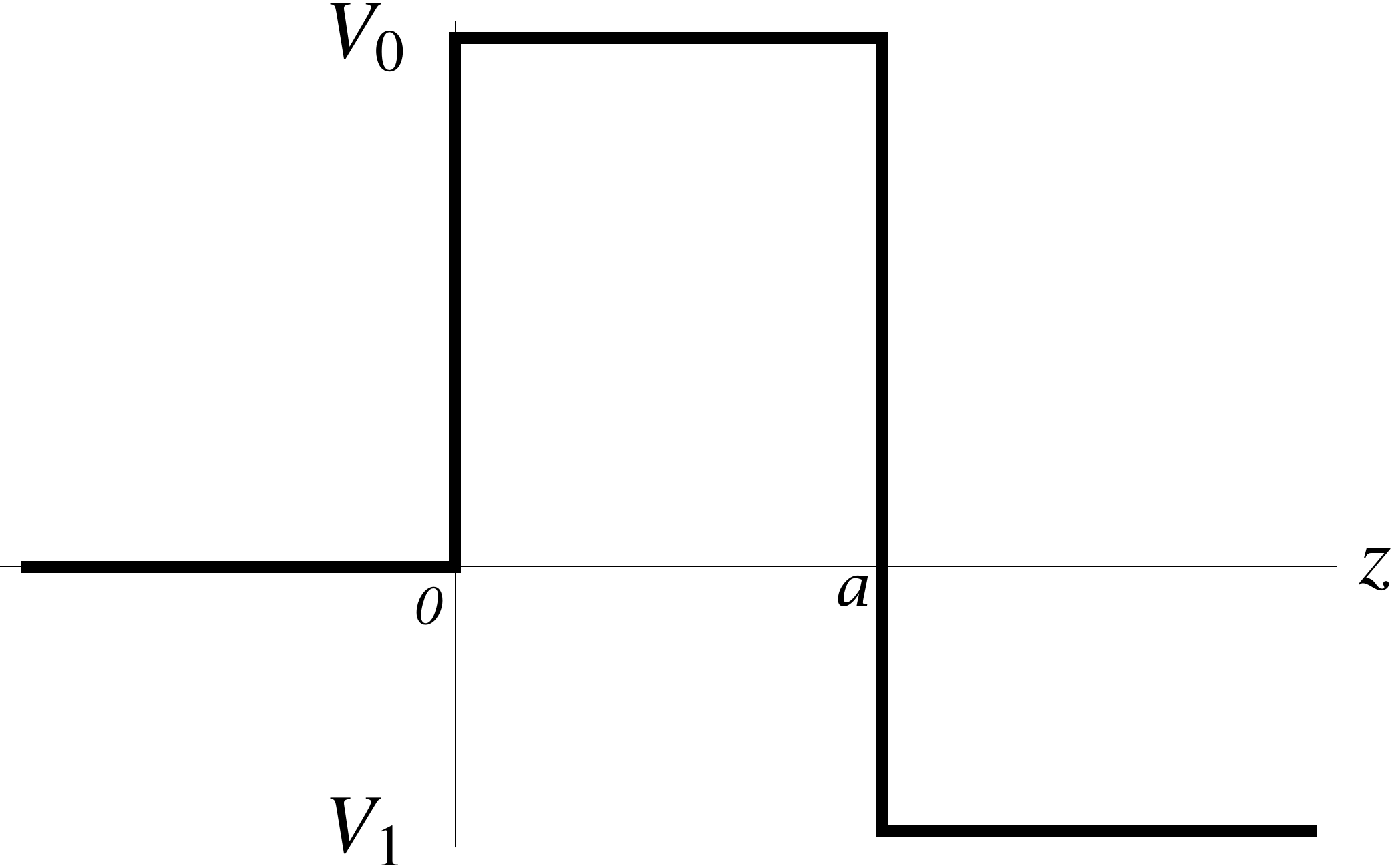}
\caption{\label{aspot} An asymmetric potential of the type considered in this work, with $\left|V_1\right|<V_0$; $V_1$ can be positive or negative.}
\end{center}
\end{figure}

Let us consider the scattering of a non-relativistic particle of mass $\mu$ and energy $E(k_1) \equiv \frac{\hbar^2 k_1^2}{2\mu}$ by an asymmetric potential of the form $V(z) = V_0 \Theta (z)\Theta(a-z) +V_1\Theta(z-a)$, where $V_0$ is assumed to be a positive constant and $V_0>|V_1|$, with $V_1$ also constant (see Fig. \ref{aspot}). We consider that the particle is incident from the left and that the SWP clock runs only while the particle is in the region $(0,a)$ -- in this case, the derivative with respect to $\mathcal{V}_m$ in (\ref{tt}) coincides with the derivative with respect to $V_0$. Also, as it is well known, for this potential the transmission and reflection coefficients are $T({\mathbf k}) = \frac{k_3}{k_1} \left| C({\mathbf k})\right|^2$ and $R({\mathbf k}) = \left| B({\mathbf k})\right|^2$, respectively; the wave numbers in the potential and transmitted regions are related to $k_1$ by $k_2 = \frac{1}{\hbar}\sqrt{2\mu\left[E(k_1) -V_0\right]}$ and $k_3 = \frac{1}{\hbar}\sqrt{2\mu\left[E(k_1) -V_1\right]}$.

The transmission and reflection amplitudes for the asymmetric potential above are easily calculated and can be found, for example, in \cite{CLi05,MPo03}. The phase $\varphi_T({\mathbf k})$ of the transmission amplitude can be obtained from $C ({\mathbf k}) = |C({\mathbf k})| e^{i[\varphi_T({\mathbf k}) - k_3 a]}$ and, for \emph{propagating modes}, it is given by
\begin{equation} \label{pt}
\varphi_T({\mathbf k}) =  \tan^{-1} \left[ \frac{(k_2^2+k_1k_3)}{k_2(k_1+k_3)}\tan(k_2a)\right]\, .
\end{equation}
Following a notation similar to that in \cite{CLi05}, the reflection amplitude can be written as $B = CGe^{ik_3a}$ and, if $\varphi_0 ({\mathbf k})$ indicates the phase of $G({\mathbf k})$, it follows that $B({\mathbf k}) = |C({\mathbf k})G({\mathbf k})| e^{i\varphi_R(k_1)}$, where the reflection phase is given by $
\varphi_R ({\mathbf k})=\varphi_T({\mathbf k})+\varphi_0({\mathbf k})$. The explicit expression for $\varphi_0$ is obtained from (\ref{pt}) by substituting $k_3\rightarrow -k_3$.

From (\ref{tt}) and (\ref{pt}) it follows that the stationary SWP transmission time for propagating energies is given by
\begin{eqnarray} \label{tct}
t_c^T ({\mathbf k}) = \frac{\mu (k_1+k_3)}{\hbar k_2} \,
\frac{ \left[k_2a(k_2^2+k_1k_3)\sec^2(k_2a)+(k_2^2-k_1k_3)\tan(k_2a)\right] }{\left[ k_2^2(k_1+k_3)^2+(k_2^2+k_1k_3)^2\tan^2 (k_2a)\right]}\, .
\end{eqnarray}
The corresponding expressions for evanescent modes can be obtained from the above equations by making the substitution $k_2 \rightarrow i q_2$, with $q_2 = \frac{1}{\hbar}\sqrt{2\mu[ V_0-E(k_1)]}$, as usual, provided that $E(k_1)>V_1$. If $V_1>0$ and $E(k_1)<V_1$ the transmission time is not defined, since there is no transmitted wave at $z>a$.

The SWP clock reflection time can be obtained by defining $t_0 ({\mathbf k}) \equiv-\hbar\frac{\partial \varphi_0}{\partial V_0}$ and noticing that
\begin{equation} \label{tcr}
t_c^R({\mathbf k})=t_c^T({\mathbf k})+t_0({\mathbf k})\, .
\end{equation}
The auxiliary time $t_0({\mathbf k})$, which can be obtained from (\ref{tct}) by making $k_3\rightarrow -k_3$, vanishes in the symmetric case and allows the separation of the contributions to $t_c^R({\mathbf k})$ due to the asymmetry of the potential (analogous to what happens for the phase time \cite{CLi05}). In addition, from (\ref{rel}) and conservation of probability, one can write the dwell time as
\begin{eqnarray} \label{td}
\tau_D({\mathbf k})&=&t_c^T({\mathbf k})+R({\mathbf k}) t_0({\mathbf k}) \nonumber \\
&=&t_c^R({\mathbf k})-T({\mathbf k}) t_0({\mathbf k}),
\end{eqnarray}
which makes clear that the dwell and SWP clock times differ only in the asymmetric case.

The SWP clock provides a ``residence'' time, i.e., it reads the time spent by particle under the potential, similarly to the dwell time. However, while the dwell time averages over the reflected and transmitted channels, eqs. (\ref{tcr})-(\ref{td}) show that \emph{the SWP clock has the advantage of distinguishing between the final channels} in the stationary case -- evidently, such distinction can only occur for asymmetric potentials, since by symmetry the two channels \emph{must} give the same results for symmetric potentials.

It is interesting to notice that a relationship can be established between the dwell time and $\nu (E)$, the average electronic density of energy per unit length \cite{GPo93,GOR95,GOC96,Ian95}. Temporarily indicating the explicit dependence of amplitudes and times in terms of the energy and making use of (\ref{rel}), such relationship can be generalized as
$$2\pi \hbar a \nu (E) = 2 T(E) t_c^T(E) +R_-(E)t_{c-}^R(E) + R_+(E)t_{c+}^R(E)\; ,$$
where the $-$ ($+$) indicates incidence from the left (right) and we used the fact that the transmission amplitude is not affected by the incidence direction. For symmetric potentials the above expression clearly reduces to $ t_{c}^{T(R)} (E) =\tau_{D } (E) = \pi \hbar a\nu (E)$.

\begin{figure}
\includegraphics[width=8.9cm,height=4.7cm]{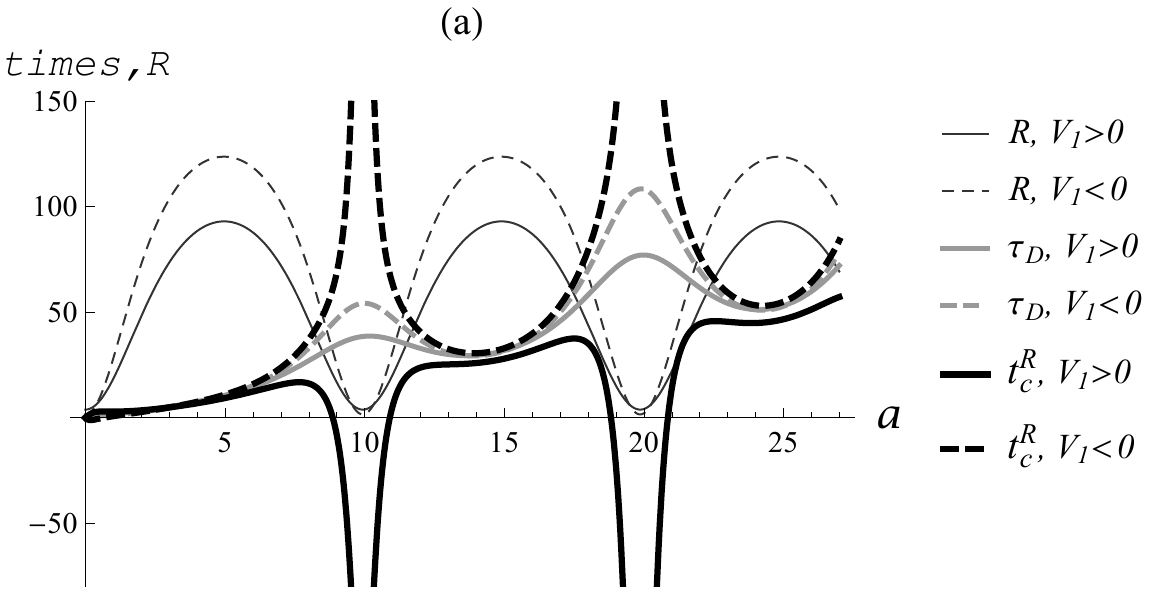}
\includegraphics[width=8.9cm,height=4.7cm]{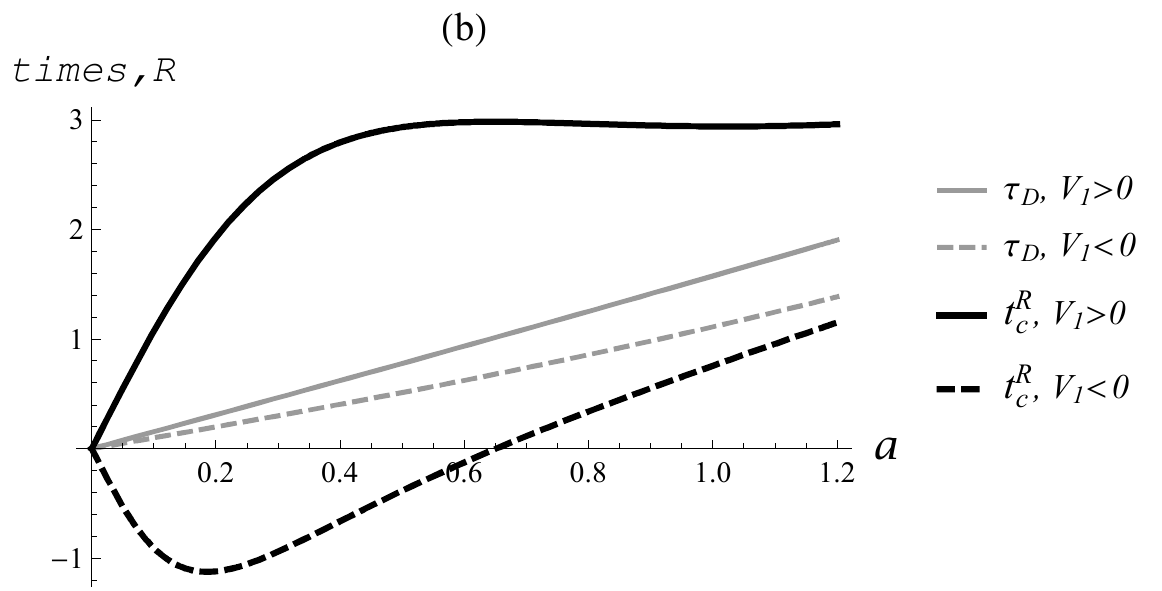}
\includegraphics[width=8.7cm,height=5.0cm]{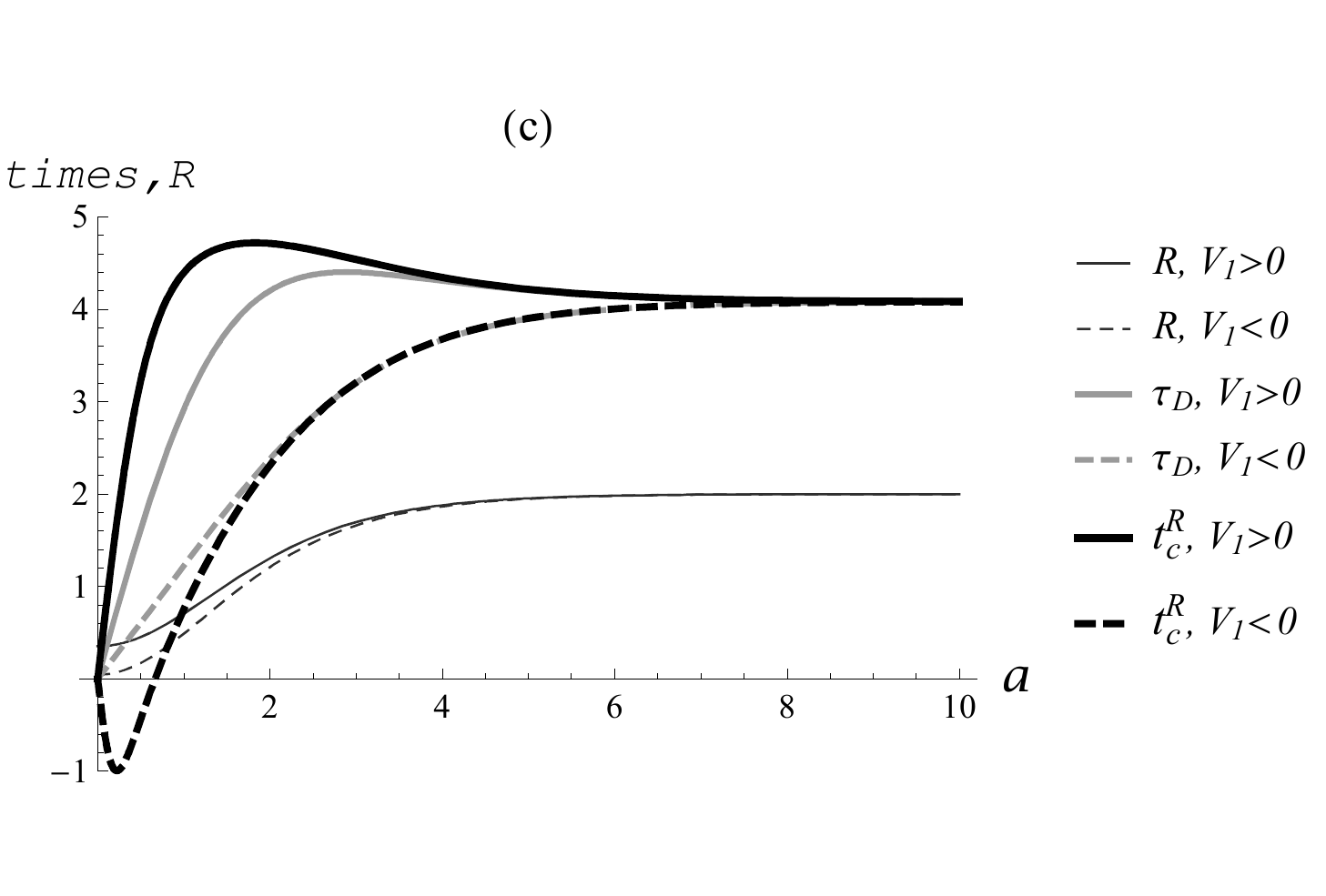}
\caption{\label{stat} Typical behavior of the stationary reflection clock time and the dwell time (thick curves) with respect to the barrier width $a$. The reflection coefficient (thin curve) is also shown in plots (a) (scaled by a factor $200$) and in plot (c) (scaled by a factor $2$). All quantities are expressed in atomic units (a.u.), $\mu = 1$ and $\hbar = 1$, with $V_0 = 0.30$ and $|V_1| = 0.15$. (a) propagating case, with $E\left(k_1\right) = 0.35$. (b) detail of plot (a) for small values of the barrier thickness. (c) evanescent case, with $E\left(k_1\right) = 0.18$.}
\end{figure}

A partial analysis of the \emph{stationary} SWP times for asymmetric potentials appears in \cite{MPo03}, which \emph{only} considers the equivalent to $t_c^T({\mathbf k})$, as one of the time scales associated with the Larmor time. Hence, before considering the average times introduced in the previous section it is important to complement those investigations by analyzing the properties of $t_c^R({\mathbf k})$.

Figure \ref{stat} displays typical plots of the stationary SWP reflection and dwell times; in addition, the coefficient of reflection $R({\mathbf k})$ is also shown. Figure \ref{stat}(a) considers a propagating energy and Fig. \ref{stat}(b) shows a detail of \ref{stat}(a) for transparent barriers. We observe that at the vicinity of the resonances, which occur at $k_2a = n\pi$ with $n$ a positive integer \cite{CLi05,MPo03}, the reflection time shows large peaks that can be positive or negative depending on the sign of $V_1$. In fact, at the resonance $t_c^R({\mathbf k})=- (2\mu k_1a/\hbar k_2^2)(V_0-V_1)/V_1$ and we see that the peak has the opposite sign of $V_1$.
Figure \ref{stat}(c) shows a plot for evanescent energies and, as expected, $t_c^R({\mathbf k})$ and $\tau_D({\mathbf k})$ saturate to the same value for opaque barriers regardless of the sign of $V_1$ [see (\ref{td})]. In the limit of transparent barriers, $k_2a \ll 1$,  the SWP reflection time behaves as $
t_c^R({\mathbf k}) \sim 2\hbar k_1a/V_1$, and it is negative for $ V_1<0$, for both propagating and evanescent modes, as it is clearly shown in Figs. \ref{stat}(b) and  \ref{stat}(c) (it is worth noticing that this is exactly the opposite of what happens with the phase time, which is negative for transparent barriers only if $V_1>0$ \cite{CLi05}). We observe that both for propagating and evanescent modes the SWP reflection time differs significantly from the dwell time only when the reflection coefficient is small [equivalently, large $T(\mathbf{k})$], \emph{i.e.,} for transparent barriers and close to resonances -- this is, of course, a direct consequence of (\ref{td}).

\begin{figure}
\includegraphics[width=8.9cm,height=5cm]{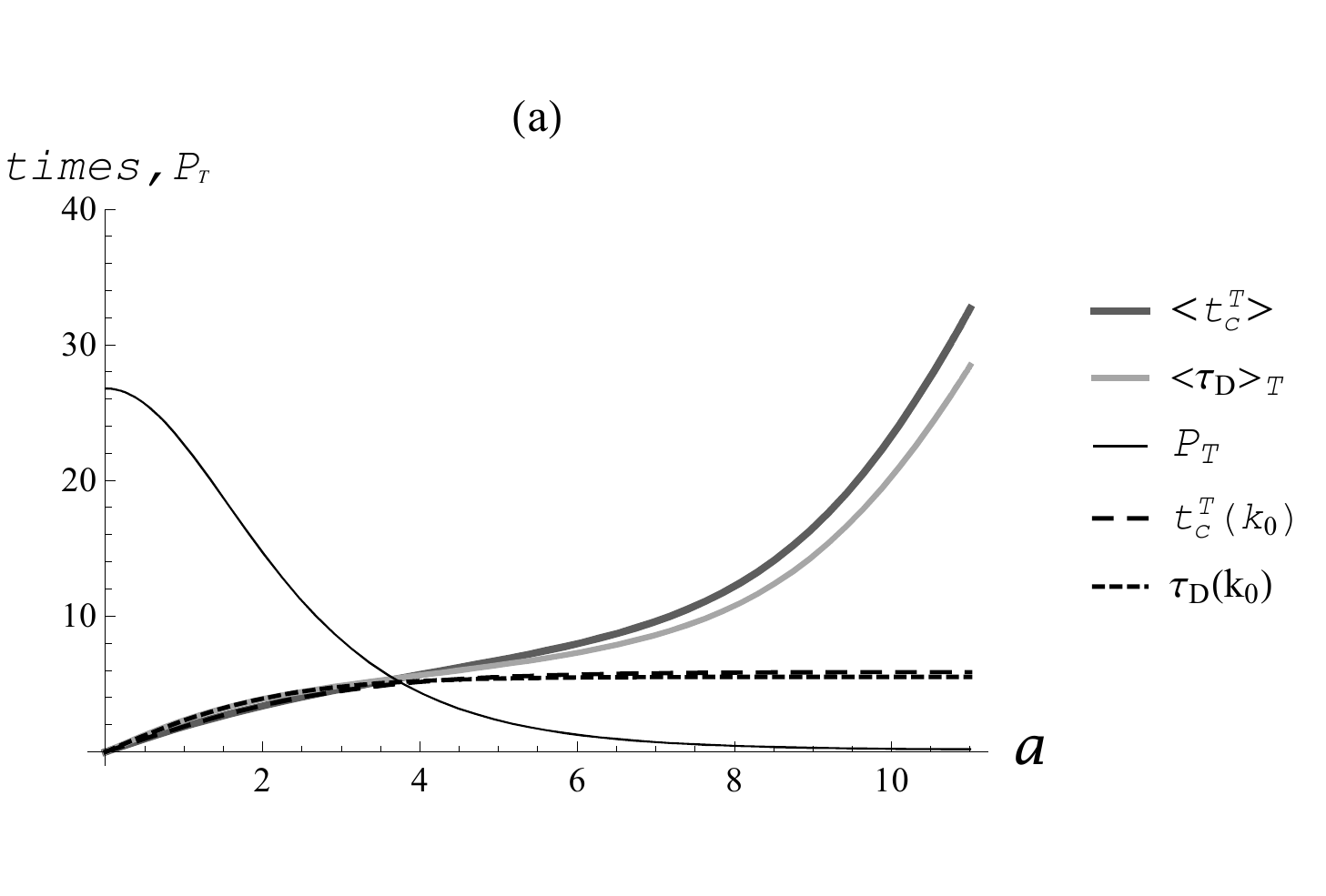}
\includegraphics[width=8.9cm,height=5cm]{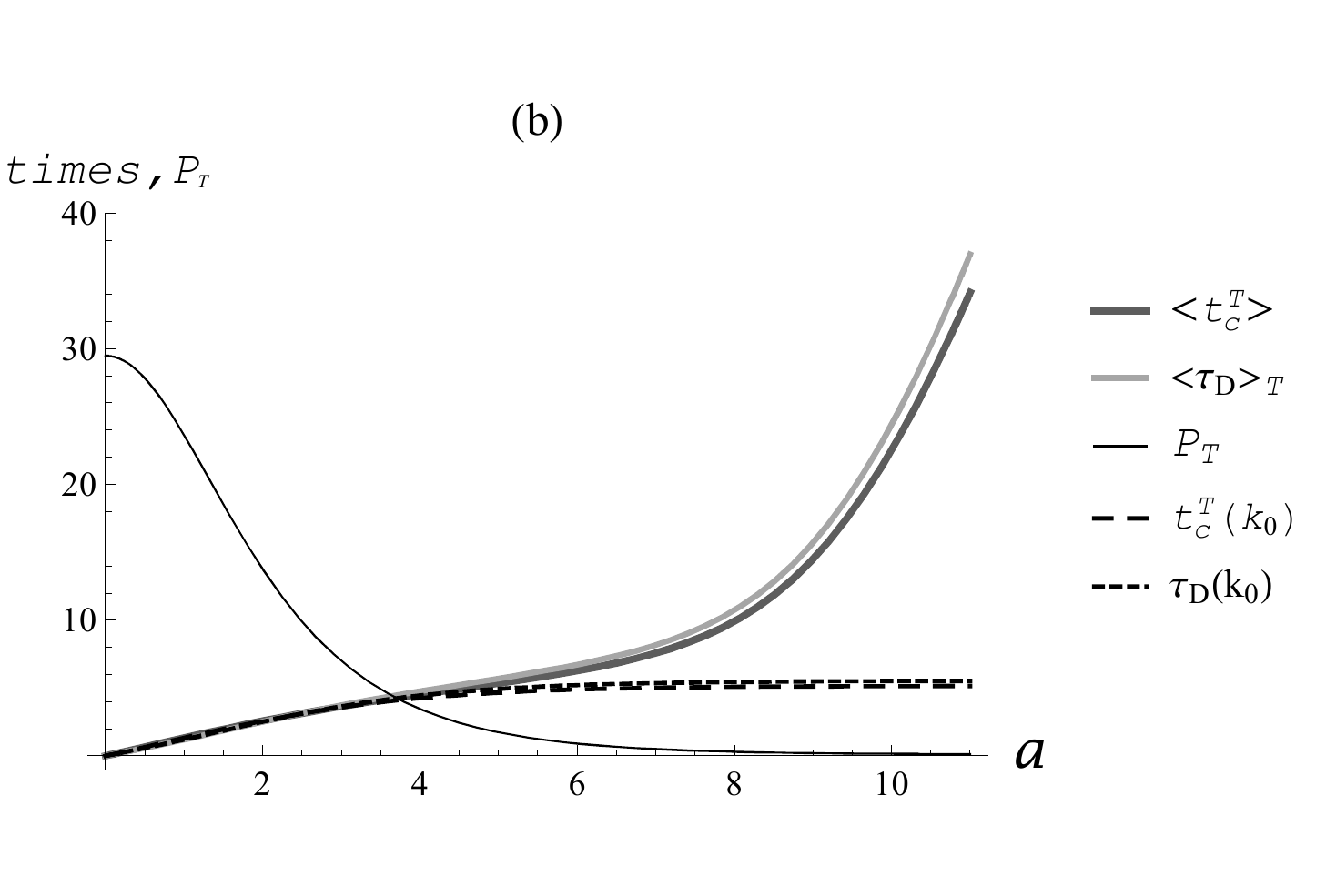}
\includegraphics[width=8.9cm,height=5cm]{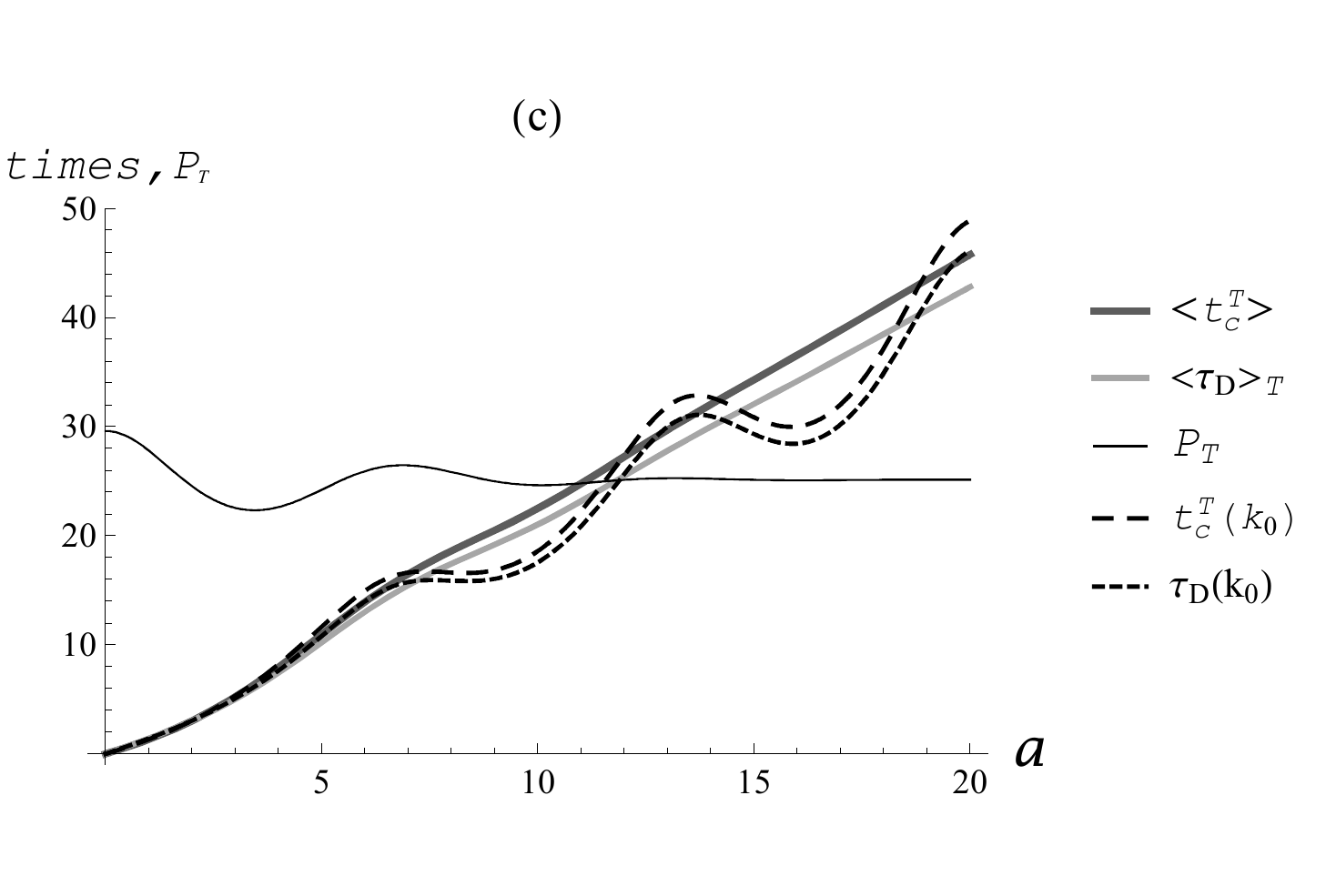}
\includegraphics[width=8.9cm,height=5cm]{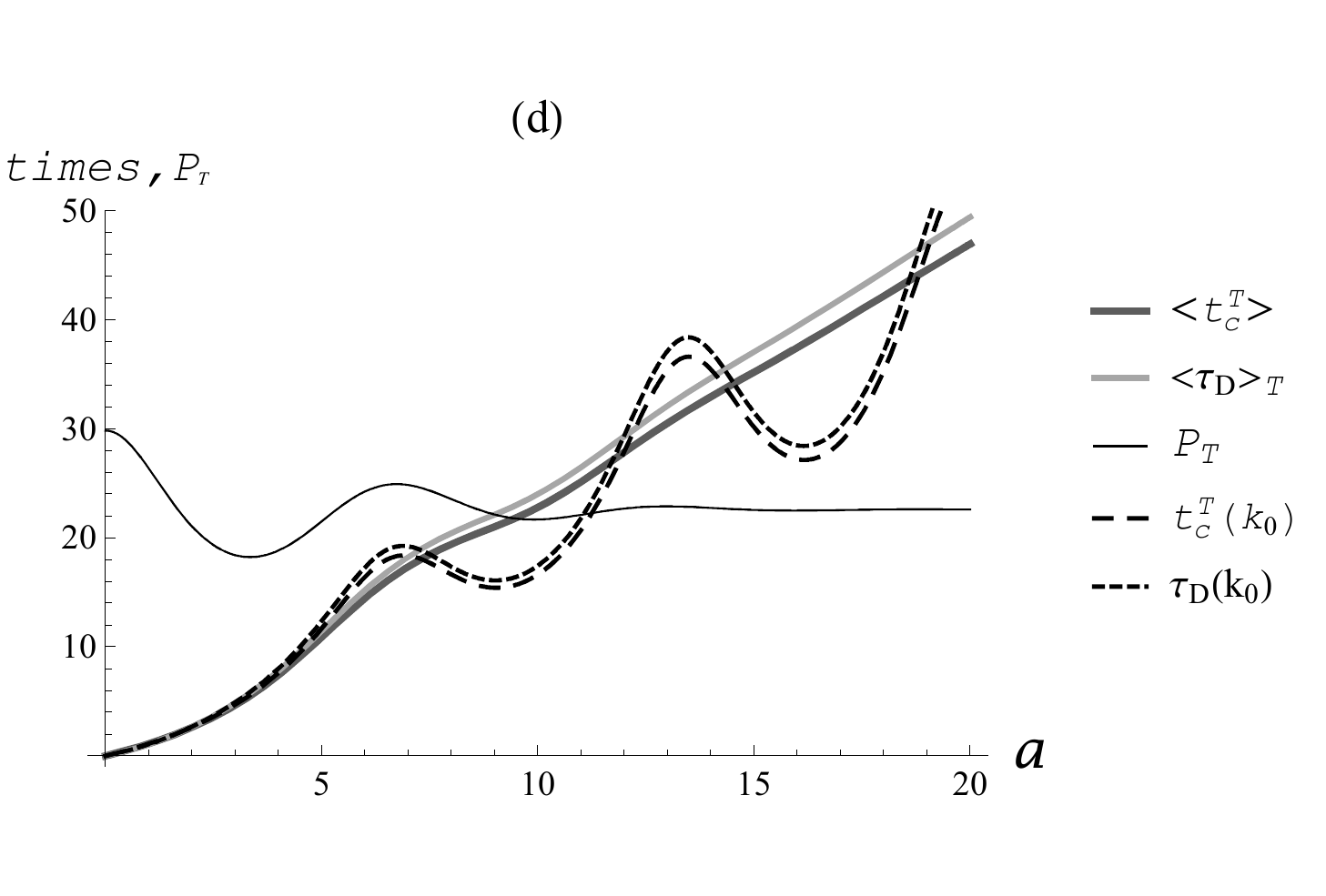}
\caption{\label{avT} Behavior of the average transmission clock time, $\langle t_c^T\rangle$, and the average dwell time, $\langle\tau_D\rangle_T$, with respect to the barrier width $a$. The stationary times $t_c^T\left(k_0\right)$ and $\tau_D\left(k_0\right)$, corresponding to the central wave number $k_0$, and the total probability of transmission of the wave packet ($P_T$; scaled by a factor $30$) are also shown. All quantities are expressed in atomic units: $\sigma = 10$, $z_0 =-8\sigma $, $V_0 = 0.30$ and $|V_1| = 0.15$. (a) $E(k_0) = 0.22$, $V_1>0$. (b) $E(k_0) = 0.22$, $V_1<0$. (c) $E(k_0) = 0.41$, $V_1>0$. (d) $E(k_0) = 0.41$, $V_1<0$.}
\end{figure}

Let us turn to the main subject of this work and focus on the \emph{average} times and consider a right-moving Gaussian \emph{wave packet} centered around a tunneling wave number $k_0 > 0$ with spatial width characterized by $\sigma$, and which at $t = 0$ is spatially centered around $z_0 < 0$. The $k$-space representation of such wave packet is given by $
A(k_1) = ( 2\sigma^2/\pi )^{\frac{1}{4}} \exp \left\{ -iz_0(k_1-k_0) -\sigma^2(k_1-k_0)^2 \right\}
$.

In Fig. \ref{avT} we compare the behavior of the average clock transmission time, $\langle t_c^T\rangle$, and the dwell time averaged over the transmitted sub-ensemble,  $\langle\tau_D\rangle_T$. Figures \ref{avT}(a) and \ref{avT}(b) refer to a wave packet with central energy $E_0$ in the tunneling region. The corresponding stationary times evaluated at the central wave number, $t_c^T\left(k_0\right)$ and $\tau_D\left(k_0\right)$, are shown to provide a reference as to the saturation width and we also show the total transmission probability $P_T = \int \frac{dk_1}{\sqrt{2 \pi}} \left| A(k_1)\right|^2 T({\mathbf k})$ associated with the wave packet. It can be seen that $\langle t_c^T\rangle$ and $\langle \tau_D\rangle_T$ behave qualitatively in very similar ways, with small quantitative differences, especially for thin barriers (in which case, their behavior is also very close to the corresponding stationary times). It is observed that for a finite dispersion $\sigma$ neither $\langle t_c^T\rangle$ nor $\langle \tau_D\rangle_T$ saturate with the barrier width ``$a$'' but, in fact, both average times tend to behave linearly with the barrier width in the extreme opaque regime -- however, both times show a residual consequence of the Hartman effect in the form of a very slow growth behavior for intermediate regions of the barrier width. It is worth noticing that the larger $\sigma$ the larger will be this intermediate region (for $\sigma \rightarrow \infty$ one recovers the saturated result of the stationary case). The above results are consistent with those obtained in \cite{LMN11}, which analyzed $\langle t_c^T\rangle$ for \emph{symmetric} potentials (also see \cite{BSM94}, which considered an analogous average for the phase time). It should be noticed that the slow growth region, which starts around the saturation width for the stationary time, is characterized by relatively small probabilities for the particle transmission. Figures \ref{avT}(c) and \ref{avT}(d) correspond to an initial wave packet centered  on a propagating energy. Again, we observe that $\langle t_c^T\rangle$ and $\langle\tau_D\rangle_T$ behave similarly over the whole range of barrier widths and tend to coincide for thin barriers.

The quantitative difference between $\langle \tau_D\rangle_T$ and $\langle t_c^T\rangle$ is given by the average of $R({\mathbf k})t_0({\mathbf k})$ over the transmitted ensemble, since from (\ref{av-d}) and (\ref{td}) it follows that $\langle \tau_D\rangle_T = \langle t_c^T\rangle + \langle Rt_0\rangle_T$. As a consequence of the product $R({\mathbf k})T({\mathbf k})$ in the numerator of the average $\langle Rt_0\rangle_T$, the differences between $\langle \tau_D\rangle_T$ and $\langle t_c^T\rangle$ tend to be more pronounced in the extreme opaque region, when it is safe to assume $R({\mathbf k})\simeq 1$. Then, in the extreme opaque limit $\langle t_c^T\rangle > \langle \tau_D\rangle_T$ for $V_1 >0$ due to the fact that $\langle Rt_0\rangle_T$ becomes negative in this limit -- see Figs. \ref{avT}(a) and \ref{avT}(c). For $V_1 < 0$ the behavior of $\langle Rt_0\rangle_T$ reverses and $\langle t_c^T\rangle < \langle \tau_D\rangle_T$ in the extreme opaque limit, as can be observed in Figs. \ref{avT}(b) and \ref{avT}(d). For transparent barriers $\langle \tau_D\rangle_T$ is slightly larger (smaller) than $\langle t_c^T\rangle$ for $V_1 > 0$ ($V_1 <0$).

For $|V_1| \ll V_0$ and $|V_1| \ll E_0$ the sign of $V_1$ has little influence and the averages $\langle t_c^T\rangle$ and $\langle \tau_D\rangle_T$ are virtually identical, as one would expect from the fact that in the symmetric case the SWP transmission time and the dwell time coincide [see eq. (\ref{td}) and notice that $t_0({\mathbf k}) \rightarrow 0$ for symmetric potentials].

\begin{figure}
\includegraphics[width=8.9cm,height=4.7cm]{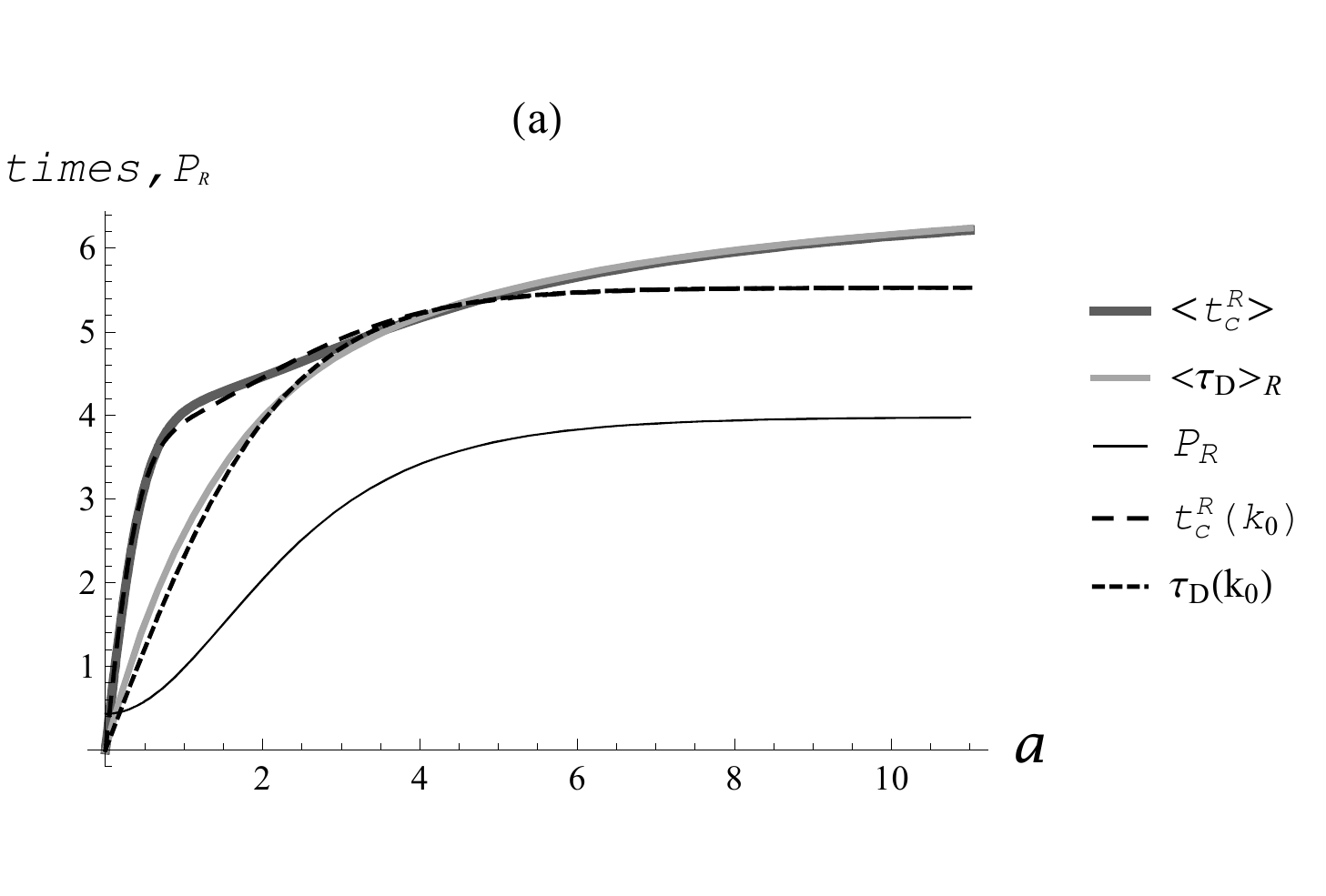}
\includegraphics[width=8.9cm,height=4.7cm]{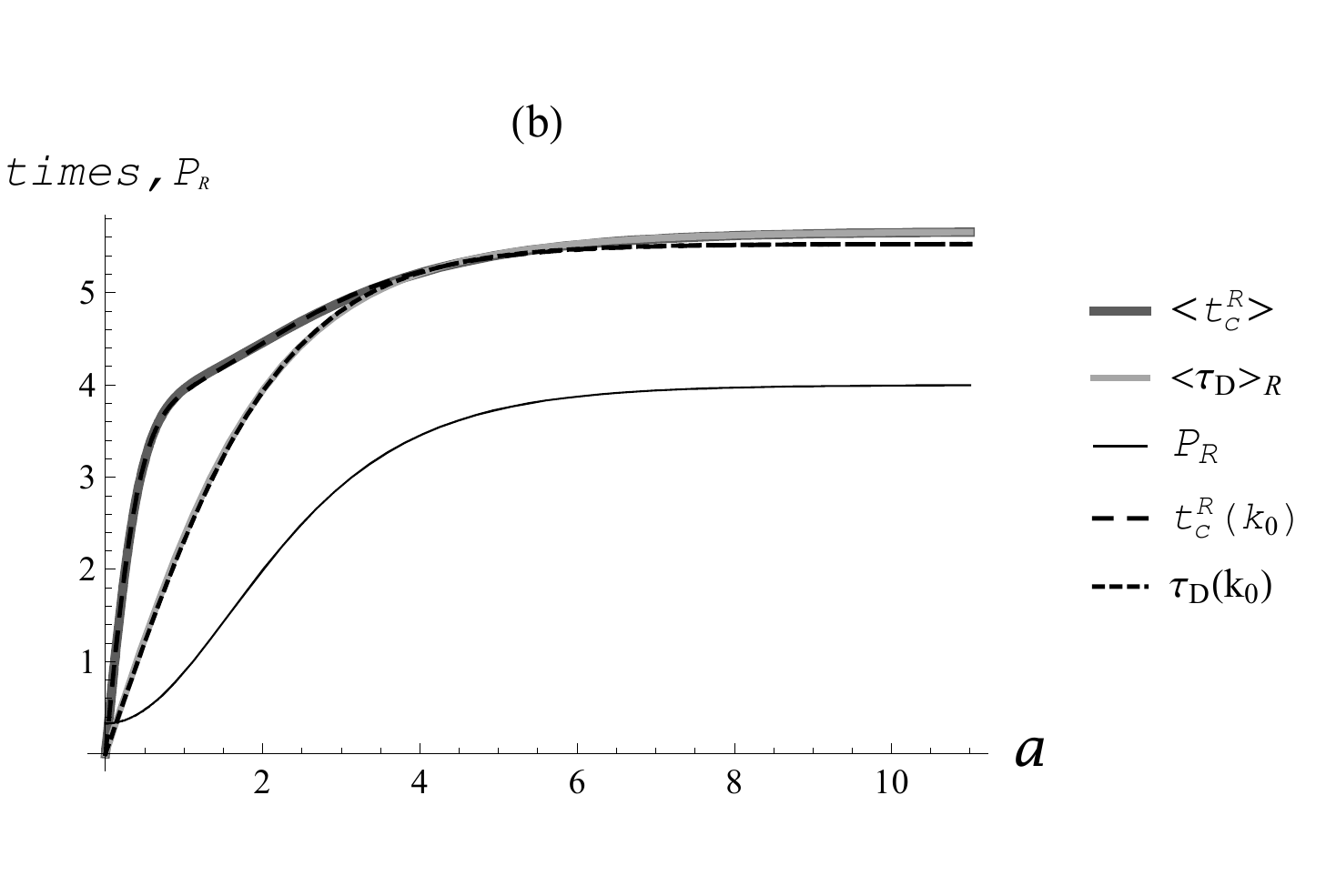}
\includegraphics[width=8.8cm,height=4.2cm]{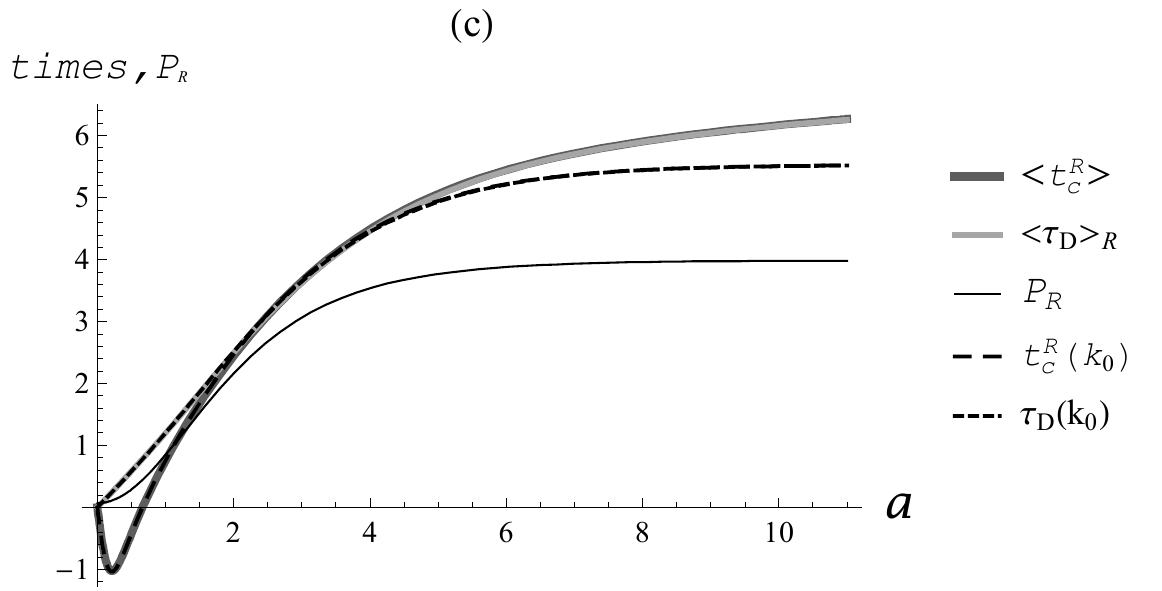}
\includegraphics[width=8.8cm,height=4.2cm]{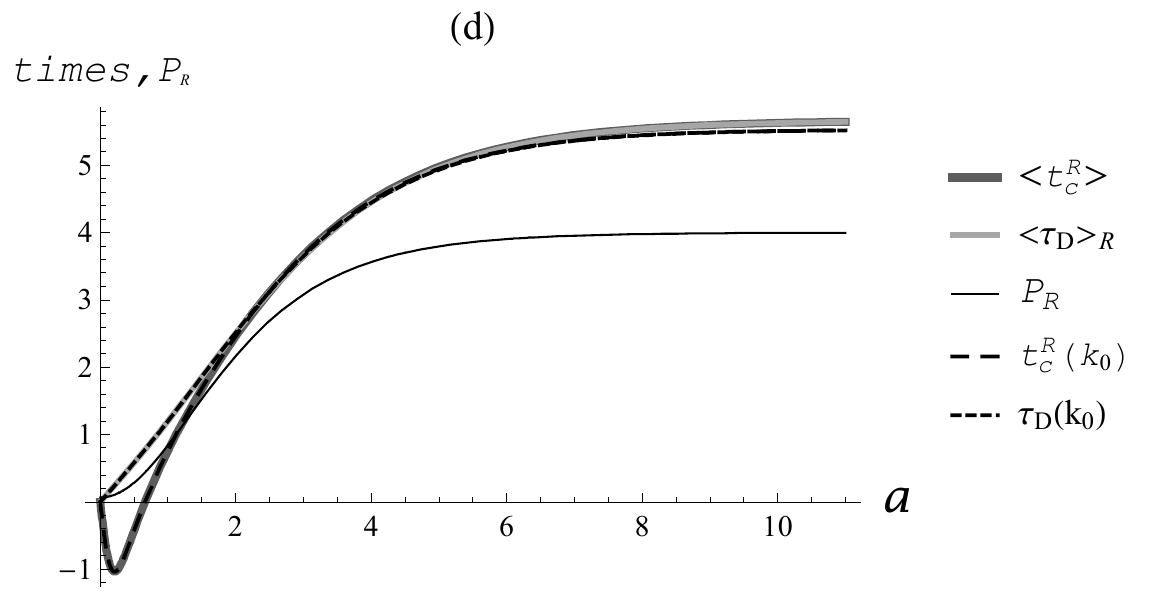}
\includegraphics[width=8.7cm,height=4.7cm]{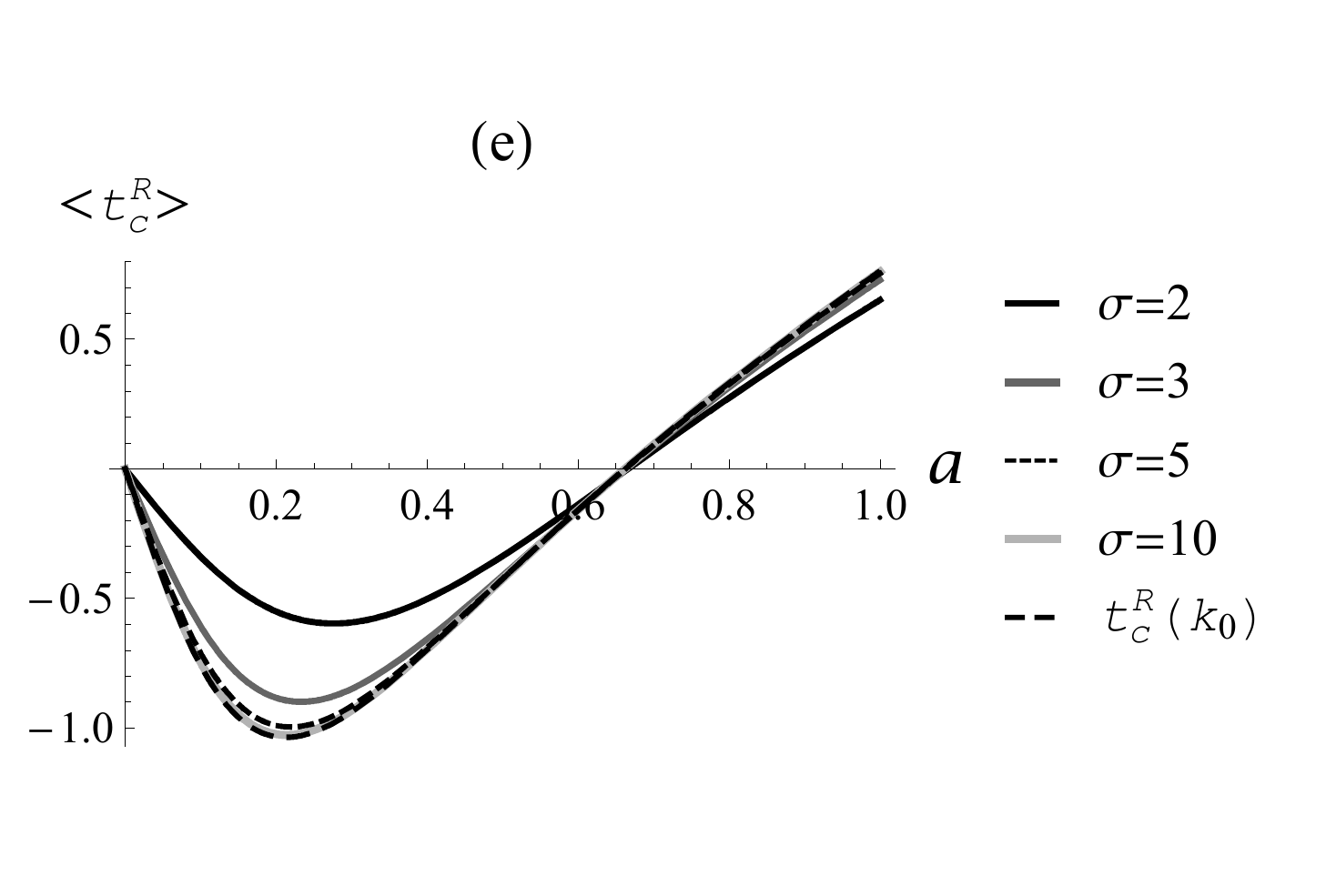}
\caption{\label{avRtun} Behavior of $\langle t_c^R\rangle$ and  $\langle \tau_D\rangle_R$ with respect to the barrier width $a$ for  an initial wave packet centered  on a tunneling energy $E\left(k_0\right)$. Except for the dispersion $\sigma$, all the parameters are the same as  in Figs. \ref{avT}(a) and \ref{avT}(b). (a) $\sigma=10$, $V_1>0$. (b) $\sigma=20$, $V_1>0$. (c) $\sigma=10$, $V_1<0$. (d) $\sigma=20$, $V_1<0$.  (e) detail of the behavior of $\langle t_c^R\rangle$ for thin barriers and $V_1<0$, for several values of the dispersion $\sigma$. In  (a)-(d) we  also plotted the total probability $P_R$ for the particle reflection (scaled by a factor $4$) and the stationary times $t_c^R\left(k_0\right)$ and $\tau_D\left(k_0\right)$.}
\end{figure}

Figure \ref{avRtun} displays typical plots for the behavior of $\langle t_c^R\rangle$ and $\langle \tau_D\rangle_R$ with respect to the barrier width $a$, for an initial wave packet peaked on a tunneling energy [the central energy $E\left(k_0\right)$ is such that $V_0>E\left(k_0\right)>V_1$].  Figures \ref{avRtun}(a) and \ref{avRtun}(b) correspond to $V_1>0$, and in this case we observe that both $\langle t_c^R\rangle$ and $\langle \tau_D\rangle_R$ are non-negative, as expected since they consist of an average over non-negative stationary times; for thin barriers both these average times are close to the respective stationary times  evaluated at the central wave number, because in this case the relative wave number composition of the reflected wave packet does not change significantly.  For opaque barriers it follows from (\ref{td}) that the stationary times coincide ($T\sim 0$), and as a consequence $\langle t_c^R\rangle$ and $\langle \tau_D\rangle_R$ also coincide. Figures \ref{avRtun}(c) and \ref{avRtun}(d) assume $V_1<0$,  and their most striking feature is that $\langle t_c^R\rangle$ in general assumes negative values for transparent barriers. This follows from the fact that, as mentioned above, for thin barriers the average time $\langle t_c^R\rangle$ closely resembles the stationary time, $t_c^R (\mathbf{k})$, which is negative for very thin barriers whenever $V_1<0$. Notice that the region in which the average clock reflection time can be negative is characterized by relatively small probabilities of reflection.

Figure \ref{avRtun} also illustrates the dependence of $\langle t_c^R\rangle$ and $\langle \tau_D\rangle_R$ with the spatial dispersion $\sigma$. It is seen that for initial wave packets sharply peaked at a tunneling mode $k_0$ [large $\sigma$; see Figs. \ref{avRtun}(b) and \ref{avRtun}(d)] both average times tend to saturate to the same value as the corresponding stationary times. On the other hand, for smaller $\sigma$'s the contributions of the over-the-barrier modes become important, leading to the slow growth of both these average times [see Figs. \ref{avRtun}(a) and \ref{avRtun}(c)]. In addition, Fig. \ref{avRtun}(e) shows that for transparent barriers, with $V_1<0$, $\langle t_c^R\rangle$ is little affected by the wave packet spatial dispersion and it is very close to the corresponding stationary time for $\sigma$ between $3$ a.u. and $10$ a.u. Significant deviations from $t_c^R(\mathbf{k})$ occur only for very small values of $\sigma$ but in this case negative wave numbers begin to play an important role in the wave packet composition invalidating the derivation of $\langle t_c^R\rangle$, which assumes only positive wave numbers \cite{LMN11}. Finally, from Figs. \ref{avRtun}(a)-(d) we see that when the \emph{total} probability of reflection is high ($P_R \simeq 1$) the average times $\langle t_c^R\rangle$ and $\langle \tau_D\rangle_R$ tend to coincide.

\begin{figure}
\includegraphics[width=8.9cm,height=4.7cm]{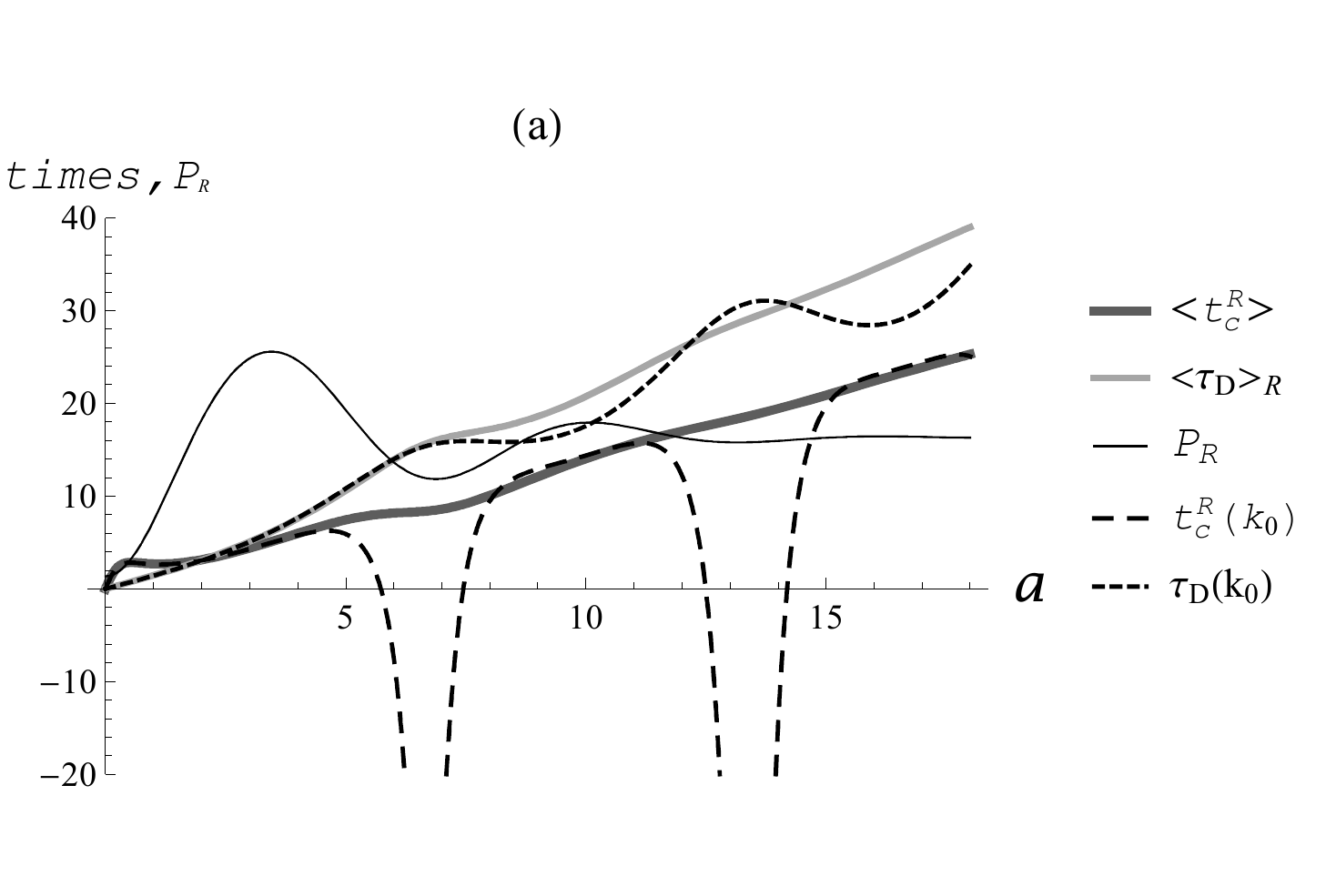}
\includegraphics[width=8.9cm,height=4.7cm]{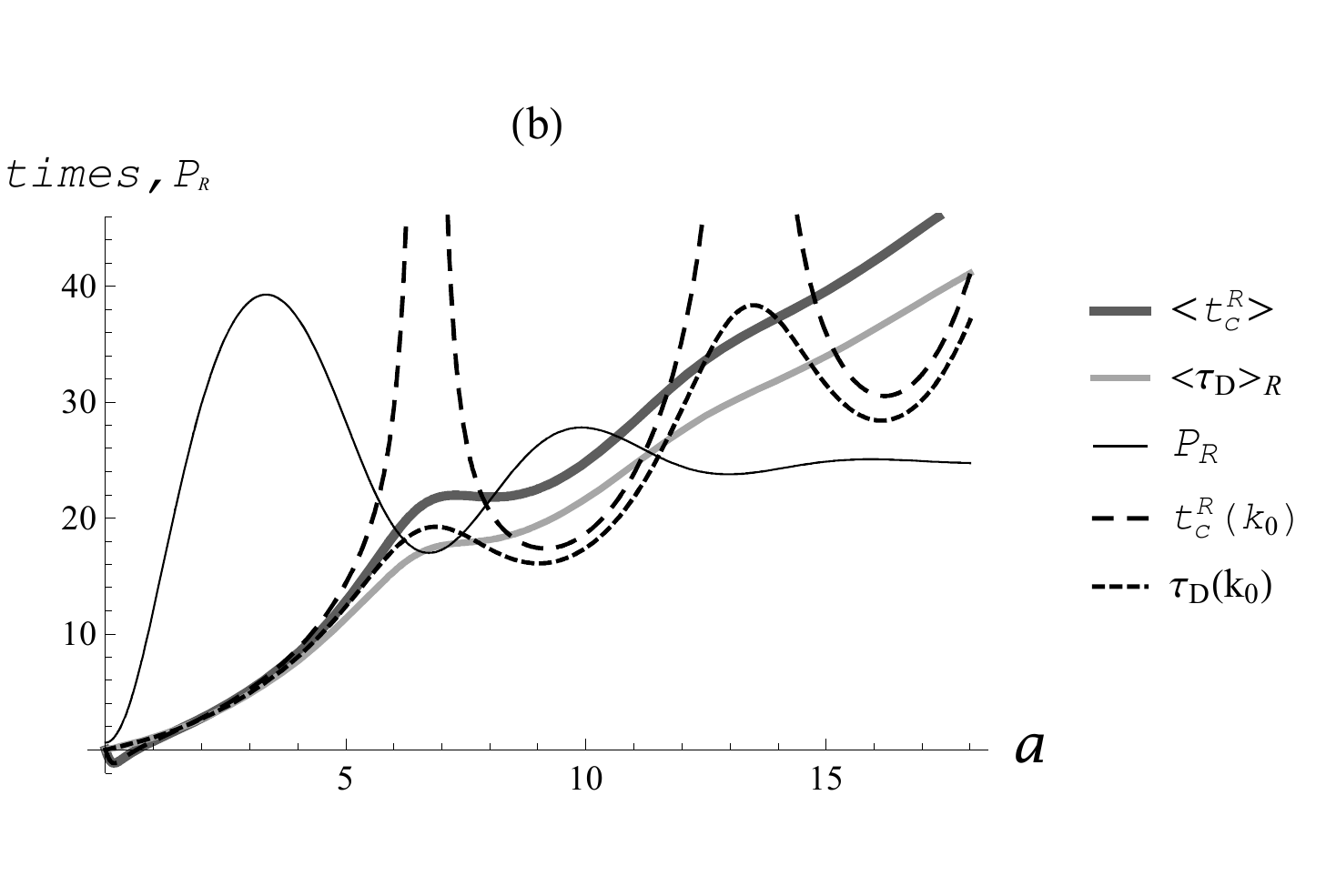}
\caption{\label{avRover} Typical behavior of the average times $\langle t_c^R\rangle$ and $\langle \tau_D\rangle_R$ \emph{versus} the barrier width $a$, for an initial wave packet centered on a propagating energy. The corresponding stationary times and the total probability of reflection ($P_R$; scaled by a factor $100$) are also plotted for reference. The parameters in (a) and (b) are the same as in Figs. \ref{avT}(c) and \ref{avT}(d), respectively. }
\end{figure}

Figure \ref{avRover} shows the typical behavior of the average times $\langle t_c^R\rangle$ and $\langle \tau_D\rangle_R$ with respect to the barrier width $a$, for an initial wave packet centered on a propagating energy, $E\left(k_0\right)$. A finite dispersion $\sigma$ has the effect of smoothing out the peaks (negative or positive) of $\langle t_c^R\rangle$ at resonances (with respect to $k_0$), in such a way that for thick barriers the qualitative behavior of $\langle t_c^R\rangle$ and $\langle \tau_D\rangle_R$ is similar [the values of these two averages tend to be closer as the probability of reflection increases, as expected from (\ref{td})]. For wave packets sufficiently localized in the configuration space $\langle t_c^R\rangle$ is always positive for thick barriers. In the region of very thin barriers both $\langle t_c^R\rangle$ and $\langle \tau_D\rangle_R$ behave similarly to the corresponding stationary times, and this implies that $\langle t_c^R\rangle$ assumes negative values for a sufficiently thin barrier when  $V_1 <0$ [recall Figs. \ref{stat}(a)-(b)]; again, the total probability of reflection in this region is relatively small ($\lesssim 8\%$).


\section{Concluding remarks}
\label{concl}

We considered the one dimensional scattering of a wave packet through a static asymmetric potential and investigated the behavior of the average transmission and reflection SWP clock times, which were compared to the behavior of the dwell time averaged over the transmitted and reflected sub-ensembles, respectively.

We evaluated the average transmission time for an initial wave packet centered on a tunneling energy and verified that it does not saturate in the opaque regime; in fact, $\langle t_c^T\rangle$  increases linearly with the barrier width in the extreme opaque limit, a result similar to the one obtained for symmetric potentials. We observed that $\langle t_c^T\rangle$ behaves qualitatively in a very similar way to $\langle \tau_D\rangle_T$, the dwell time averaged over the transmitted sub-ensemble, for wave packets centered on both evanescent and propagating energies -- both these average times provide good scales for the particle transmission across an asymmetric barrier.

The clock's reflection time was also investigated. In the stationary case it allows for negative values in two situations: for $V_1<0$ and very thin barriers (in both the propagating and evanescent cases), and for $V_1>0$ at resonances (in the propagating case). This is in contrast with the stationary dwell time, which is always non-negative. We considered the average time $\langle t_c^R\rangle$ for a localized wave packet and concluded that the negative values persist in the average time for thin barriers, because in this case the average behaves very similarly to the stationary time (such a result is associated with a relatively small probability of reflection). On the other hand, $\langle t_c^R\rangle$ behaves well at the vicinities of resonances of its central component; in fact, the average has the effect of smoothing out the (positive or negative) peaks found in the stationary case. This can be understood by observing that resonant modes have a very small probability of reflection, and then do not contribute significantly to such an average.

In general, for non-transparent barriers, the average clock time of reflection behaves in a (qualitatively) similar way to the dwell time averaged over the reflected sub-ensemble  (although, of course, there are quantitative differences which could only be settled one way or the other on an experimental basis). Thus, based on the properties described in the previous section, both $\langle t_c^R\rangle$ and $\langle \tau_D\rangle_R$ are potentially good time scales to describe the reflection of well localized particles by asymmetric barriers. The fact that $\langle t_c^R\rangle$ may lead to negative values for transparent barriers, depending on the parameters, does not discard it as a good scale since these values are in general associated with relatively low probabilities and may be explained using the weak measurement theory \cite{AAV88,AhV90,Ste95} -- it is well known that the \emph{stationary} SWP times are weak values \cite{Dav05,Ian96}; however, an in-depth study of the SWP clock \emph{for wave packets} in the context of the weak measurement theory is still necessary to clarify this point. Finally, $\langle t_c^{R(T)}\rangle$ has the advantage (with respect to $\langle \tau_D\rangle_{R(T)}$) that in the stationary limit the SWP clock times do distinguish between the reflected and transmitted channels, while the dwell time is an average over these channels.

\begin{acknowledgement}
The authors would like to thank two anonymous referees for several references and suggestions to improve the manuscript. This work was partially supported by NASA Minnesota Space Grant Consortium (B.A.F. and L.A.M.) and NSF STEP grant $\# 0969568$ (L.A.M).
\end{acknowledgement}



\end{document}